\RequirePackage{etex} 
\documentclass[a4paper,UKenglish,cleveref, autoref, thm-restate]{lipics-v2021}

\usepackage{xcolor}
\usepackage{tikz}
\usetikzlibrary{arrows.meta, matrix}

\usepackage{algorithm2e}
\usepackage[utf8]{inputenc}
\usepackage{inputenc}
\usepackage{amsmath}
\usepackage{amssymb}
\usepackage{ upgreek }
\usepackage{graphicx}
\usepackage{layout}
\usepackage{dsfont}
\usepackage{bbold}
\usepackage{xspace}
\usepackage{multicol}
\usepackage{amsthm}
\usepackage{mathrsfs}
\usepackage{tikz}
\usepackage{tikz-network}
\usepackage[toc,page]{appendix}
\usepackage{todonotes}
\usepackage{complexity}
\usepackage{comment}
\usepackage{subcaption}
\usepackage{amsmath,amsfonts, amssymb, bm}
\usepackage{xspace}
\usepackage[singlelinecheck=false]{caption}

\pdfoutput=1 
\hideLIPIcs  

\tikzstyle{v}=[circle,inner sep=0, minimum size =6 pt, line width = 1pt, draw=black, fill=black, text= white]
\tikzstyle{inv2}=[circle,inner sep=0, minimum size = 18 pt, line width = 0pt, draw=white, fill=white, text= black]
\tikzstyle{R}=[circle,inner sep=0, minimum size =7 pt, line width = 1 pt, draw=red, fill = red]
\tikzstyle{inv}=[circle,inner sep=0, minimum size = 0 pt, line width = 0pt, draw=white, fill=white, text= black]
\tikzstyle{B}=[circle,inner sep=0, minimum size =7 pt, line width = 1 pt, draw=blue, fill = blue]

\title{On the complexity of the Maker-Breaker happy vertex game} 

\author{Mathieu Hilaire}{Univ. Bordeaux, Bordeaux INP, LaBRI UMR CNRS 5800, F-33400, Talence, France.}{mathieu.hilaire@labri.fr}{}{}

\author{Perig Montfort}{ENS de Lyon, Lyon, France.}{perig.montfort@ens-lyon.fr}{}{}

\author{Nacim Oijid}{Umeå University, Umeå, Sweden.\and \url{https://nacim-oijid.fr}}{nacim.oijid@umu.se}{https://orcid.org/0000-0001-8313-639X}{Kempe Foundation Grant No. JCSMK24-515 (Sweden)}

\authorrunning{M. Hilaire et al} 

\Copyright{Mathieu Hilaire, Nacim Oijid and Perig Montfort} 



\category{} 

\relatedversion{} 

\funding{This research was supported by the ANR project P-GASE (ANR-21-CE48-0001-01)}

\nolinenumbers 

\EventEditors{John Q. Open and Joan R. Access}
\EventNoEds{2}
\EventLongTitle{42nd Conference on Very Important Topics (CVIT 2016)}
\EventShortTitle{CVIT 2016}
\EventAcronym{CVIT}
\EventYear{2016}
\EventDate{December 24--27, 2016}
\EventLocation{Little Whinging, United Kingdom}
\EventLogo{}
\SeriesVolume{42}
\ArticleNo{23}

\newcommand{\Maker}{Maker}
\newcommand{\Breaker}{Breaker}

\newtheorem{problem}[theorem]{Problem}
\newcommand{\Input}{\textbf{Input}}

\newcommand{\Question}{\textbf{Question}}
\newcommand{\strat}{\mathcal{S}}

\begin{document}

\maketitle

\begin{abstract}

Given a $c$-colored graph $G$, a vertex $v$ of $G$ is said to be {\em happy} if it has the same color as all its neighbors. The notion of happy vertices was introduced by Zhang and Li~\cite{ZHANG2015} to compute the homophily of a graph. Eto, Fujimoto, Kiya, Matsushita, Miyano, Murao and Saitoh~\cite{Eto2025HappyVertex} introduced the Maker-Maker version of the Happy vertex game, 
where two players compete to claim more happy vertices than their opponent. 
We introduce here the Maker-Breaker happy vertex game: two players, Maker and Breaker, alternately color the vertices of a graph with their respective colors. Maker aims to maximize the number of happy vertices at the end, while Breaker aims to prevent her. 
This game is also a scoring version of the Maker-Breaker Domination game introduced by Duchene, Gledel, Parreau and Renault~\cite{duchene2020domination}, as a happy vertex corresponds exactly to a vertex that is not dominated in the domination game. Therefore, this game is a very natural game on graphs and can be studied within the scope of scoring positional games~\cite{bagan2024incidence}. We initiate here the complexity study of this game, by proving that computing its score is \PSPACE-complete on trees, \NP-hard on caterpillars, and polynomial on subdivided stars. Finally, we provide the exact value of the score on graphs of maximum degree~2, and we provide an \FPT-algorithm to compute the score on graphs of bounded neighborhood diversity. An important contribution of the paper is that, to achieve our hardness results, we introduce a new type of incidence graph called the {\em literal-clause incidence graph} for $2$-SAT formulas. We prove that \textsc{QMAX 2-SAT} remains \PSPACE-complete even if this graph is acyclic, and that \textsc{MAX 2-SAT} remains \NP-complete, even if this graph is acyclic and has maximum degree~$2$, i.e. is a union of paths. We demonstrate the importance of this contribution by proving that Incidence, the scoring positional game played on a graph is also \PSPACE-complete when restricted to forests.
\end{abstract}

\section{Introduction}

\subsection{Framework of the study}

Happy vertices have been introduced in graphs to measure their homophily: a vertex is {\em happy} if it has the same color as all its neighbors. This concept has been introduced by Zhang and Li~\cite{ZHANG2015}, and has been the subject of several studies afterwards; see for instance~\cite{agrawal2017parameterized, aravind2016linear, lewis2019finding, zhang2018improved}. In this context, Eto, Fujimoto, Kiya, Matsushita, Miyano, Murao and Saitoh have introduced the happy vertex game~\cite{Eto2025HappyVertex} as a two-player game, in which two players, Black and White, alternately color the vertices of a graph $G$ with their own color and aim to have more happy vertices of their color than their opponent. We focus here on the Maker-Breaker version of the game, in which Maker and Breaker again alternately color the vertices of a graph, Maker in red, and Breaker in blue. Maker aims to maximize the number of happy red vertices, and Breaker aims to minimize it. Hence, when all the vertices are colored, the number of red happy vertices is precisely the number of vertices not dominated by Breaker, which makes this game a scoring version of the well-known {\em Maker-Breaker domination game}.

The Maker-Breaker domination game, introduced by Duch\^ene, Gledel, Parreau and Renault~\cite{duchene2020domination} in 2020, is one of the most studied positional games played on the vertices of graphs. In this game, two players, Dominator and Staller, alternately claim the unclaimed vertices of a given graph $G$ until no vertices remain. Dominator wins if the set of vertices she has claimed is a dominating set; otherwise, i.e. if Staller has isolated one vertex by claiming it and all its neighbors, Staller wins. The algorithmic study of this game led to its \PSPACE-completeness on bipartite graphs of bounded degree or split graphs~\cite{duchene2020domination, oijid2025qbf}, and to polynomial algorithms on several classes of graphs, including paths, grids, unit interval graphs, and regular graphs~\cite{Bagan2025, Dokyeesun2025}. This game is part of the larger field of Maker-Breaker positional games, which are games played on a hypergraph with two players, Maker and Breaker, alternately claiming the vertices of the hypergraph. Maker wins if she manages to claim all the vertices of some hyperedge; otherwise, Breaker wins. Positional games were introduced by Hales and Jewett in 1963~\cite{hales1963regularity}, and proved to be \PSPACE-complete by Schaefer in 1978~\cite{schaefer1978complexity} for hypergraphs of rank~11, i.e. whose largest hyperedge has size~$11$. The study of their complexity has recently gained strong interest with the three consecutive improvements on the minimal rank for which the game remains \PSPACE-complete to rank~$6$ by Rahman and Watson~\cite{rahman20236}, to rank~$5$ by Koepke~\cite{koepke2025advances}, and to rank~$4$ by Galliot~\cite{galliot20254uniformmakerbreakermakermakergames}. On the positive side, Galliot, Gravier, and Sivignon proved that the game is solvable in polynomial time on hypergraphs of rank~$3$~\cite{galliot2025makerbreakersolvedpolynomialtime}.

Here, in the Maker-Breaker happy vertex game, the ruleset is the same as in the Maker-Breaker domination game, except that, once all the vertices have been claimed, we count the number of vertices isolated by Staller, instead of asking if one exists. This consideration makes this game in the universe of scoring positional games, introduced by Bagan, Deschamps, Duchêne, Durain, Effantin, Gledel, Oijid and Parreau~\cite{bagan2024incidence}. They proved that scoring positional games belongs to Milnor's universe, a large framework of scoring combinatorial games, which grants some structure on the game~\cite{milnor1953sums}. They also proved that computing the score of a scoring Maker-Breaker positional game is \PSPACE-complete, even restricted to hypergraphs of rank~2, i.e. graphs, and they provided several tools to handle this class of games.

\subsection{Outline of the paper}

In this paper, we first remind some useful results about Milnor's universe and the domination game in Section~\ref{sec: preliminaries}. In particular, we use Milnor's group structure to handle disconnected graphs, which can be handled through his results on sum of games, and we use the results on the domination game to focus our study on specific graph classes.

In Section~\ref{sec: hardness}, we introduce several variants \textsc{MAX 2-SAT}, together with a new representation: the literal-clause incidence graph. We prove that a quantified version of \textsc{MAX 2-SAT} is \PSPACE-complete even when we restrict it to instances for which this graph is acyclic, and that \textsc{MAX 2-SAT} is \NP-complete, even when we restrict it to instances for which this graph is acyclic and has maximum degree~2. As a corollary, we prove that Incidence, the scoring positional game played on a graph is \PSPACE-complete, even restricted to forests, answering an open question of~\cite{bagan2024incidence}. We then turn back to the Happy vertex game, and we prove that even though the Maker-Breaker Domination game can be solved in polynomial time on forests, its scoring version that we study here is already \PSPACE-complete on trees. This phenomenon leads us to the study of specific subclasses of trees, hence, we prove that the game is already \NP-hard on caterpillars (Section~\ref{sec: hardness}), but can be solved in polynomial time on subdivided stars and union of paths (Section~\ref{sec: poly}). Concerning the latter result, we can extend it to graphs of maximum degree~$2$ since isolated cycles will not change the score of the game.

Finally, in search for extending results known from scoring positional games, we adapt the so-called super-lemma to this game and manage to provide an \FPT-algorithm in Section~\ref{sec: parameterized} parameterized by the neighborhood diversity, a graph parameter introduced by Lampis~\cite{Lampis2012} that measures the variety of the neighborhoods of vertices in the graph.

\section{Preliminaries} \label{sec: preliminaries}

\subsection{Definitions and Notations}

Let $G = (V,E)$ be a graph, and let $v$ be a vertex of $G$. We denote by $N(v)$ the {\em open neighborhood} of $v$, i.e. we have $N(v) = \{u \in V(G) \mid (u,v) \in E(G)\}$. We denote by $N[v]$ the {\em closed neighborhood} of $v$, i.e. $N[v] = \{v\} \cup N(v)$. 

We denote by \textsc{SHVG} the Scoring Happy Vertex Game, and we identify it with the decision problem that takes as input a graph $G$ and an integer $s$ and returns \texttt{true} if and only if Maker has a strategy to guarantee a score of at least \( s \) on $G$. Formally, the problem is defined as follows:

\begin{problem}[\textsc{Scoring Happy Vertex Game}]\label{shvg}{}
\hspace{.2cm}

\noindent    
\Input: A graph $G$, an integer $s$, a first player $M$ or $B$.
    
\noindent    
\Question: 
Do we have $s(G) \ge s$?
\end{problem}

As usual in positional game, if the first player is not specified, we suppose that it is Maker.

A current {\em game state} is represented by the triple \( (G, M, B) \), where \( M, B  \subset V\) are the sets of vertices colored by Maker (\Maker) and Breaker (\Breaker), respectively. We also denote by \(V_F(G)=V(G)\setminus(M\cup B)\) the set of {\em free vertices} in $G$.

We denote by \( Ms(G,M,B) \) the score obtained on \( (G,M,B) \) when \Maker{} plays next, and \( Bs(G,M,B) \) when \Breaker{} plays next. Note that we will always consider that both players play optimally, i.e. this score is uniquely defined as the maximum score that Maker can ensure against all the possible strategies for Breaker. In particular, \( Ms(G, \emptyset, \emptyset) \) denotes \Maker{}'s maximal score when \Maker{} starts the game on $G$, and \( Bs(G, \emptyset, \emptyset) \) when \Breaker{} starts. If there is no possible confusion, they will be denoted by \( Ms(G) \) and \( Bs(G) \) respectively.

When the identity of the starting player is irrelevant, we denote the score by \( s(G) \), omitting the prefix \( M \) or \( B \). This allows us to state results that are valid for both \( Ms \) and \( Bs \). However, it is important to note that this notation \textbf{does not} imply that \( Ms = Bs \); it merely indicates that the statement applies to both cases independently.

\subsection{Scoring game}

In this section, we begin by introducing some basic results about the score function associated with the game. We start by recalling the inductive definition of the score in scoring games, using the standard notations of Larsson~\cite{larsson2019games}.

\begin{definition}\label{def Ms Bs}
Let \((G, M, B)\) be a game state. Let \(V_F = V \setminus (M \cup B)\) be the set of uncolored vertices in this position. Then, the scores \(Ms\) and \(Bs\) are defined inductively as follows:
\[
\begin{cases}
Ms(G, M, B) = \max\limits_{x \in V_F} Bs(G, M \cup \{x\}, B), \\
Bs(G, M, B) = \min\limits_{x \in V_F} Ms(G, M, B \cup \{x\}).
\end{cases}
\]
If all vertices have been colored, i.e., \(M \cup B = V\), the score is evaluated by:
\[
s(G, M, B) = \left| \left\{ v \in V(G)\,\middle|\, N[v] \subseteq M \right\} \right|
\]
\end{definition}

We also recall that Milnor's universe is the set of games that are {\em dicotic} and {\em nonzugzwang}, i.e., the set of games in which, if a player can move, his opponent also can, and in which the players have no interest in passing their turn. Being a scoring positional game, the happy vertex game belongs to Milnor's universe~\cite{bagan2024incidence}, which implies the following inequality:

\begin{lemma}\label{relation Bs Ms}
For any game state \((G, M, B)\), we have:
\[
Bs(G, M, B) \leq Ms(G, M, B).
\]
\end{lemma}

Within Milnor's universe, we define the union between two games \(G\) and \(H\), and we note $G \cup H$, the game in which, at each turn, a player may choose to move either in \(G\) or in \(H\). Considering positional games played on graphs, this consists in considering that playing on a disconnected graph consists in playing on the sum of its connected components.

We also define the notion of equivalence as follows:

\begin{definition}\label{equiv}
    Two scoring games \(G_1\) and \(G_2\) are equivalent, and we note $G_1 = G_2$, if for every game \(G\), the scores satisfy
    \[
    Ms(G \cup G_1) = Ms(G \cup G_2) \quad \text{and} \quad Bs(G \cup G_1) = Bs(G \cup G_2).
    \]
\end{definition}

Moreover, the group structure in Milnor's universe makes it possible to bound the score on a union of games knowing the score of its terms. Namely, we have:

\begin{theorem}[Milnor~\cite{milnor1953sums}]\label{bound union milnor}
    Let \((G_1, M_1, B_1)\) and \((G_2, M_2, B_2)\) be two positions in the Happy vertex game. We have:
  
    \begin{align*}       
    Bs(G_1, M_1, B_1)+Bs(G_2, M_2, B_2)\leq Bs(G_1 \cup G_2, M_1 \cup M_2, B_1 \cup B_2) \\
    \leq \min \left( Ms(G_1, M_1, B_1) + Bs(G_2, M_2, B_2), Bs(G_1, M_1, B_1) + Ms(G_2, M_2, B_2)  \right )
    \end{align*}

\begin{align*}
    \max \left ( Ms(G_1, M_1, B_1) + Bs(G_2, M_2, B_2),  Bs(G_1, M_1, B_1) + Ms(G_2, M_2, B_2) \right ) \leq \\
    Ms(G_1 \cup G_2, M_1 \cup M_2, B_1 \cup B_2) \leq Ms(G_1, M_1, B_1) + Ms(G_2, M_2, B_2)
    \end{align*}

\end{theorem}

In particular, if $Ms(G_1) = Bs(G_1)$, since $Ms \ge Bs$, the left and the right parts of the inequalities are equal, we have the following result:

\begin{theorem}[Milnor 1953~\cite{milnor1953sums}]\label{adding game of fixed score}
    Let $(G,M,B)$ be a game position such that $Ms(G,M,B) = Bs(G,M,B) = s$. Then, we have for any position $(G',M',B')$:
$$ s(G'\cup G, M'\cup M, B' \cup B) = s(G',M',B') + s$$
\end{theorem}

This result takes place in the more general framework of {\em temperature theory}, by stating that if a game $G$ satisfies $Ms(G) = Bs(G)$, then its temperature is $0$. However, the framework of temperature theory will not be required throughout this article. We refer the reader to Hanner's results~\cite{Hanner1959MeanPO} and to an article by Duchêne, Oijid, and Parreau~\cite{DUCHENE2024114274} that summarizes these results for a general overview of this framework.

\subsection{Results inherited from the Maker-Breaker domination game}

The Maker-Breaker domination game can result in three different outcomes: \(\mathcal{D}\), \(\mathcal{S}\), and \(\mathcal{N}\). The outcome \(\mathcal{D}\) indicates that Dominator wins going second. The outcome \(\mathcal{S}\) means that Staller wins going second. Finally, the outcome \(\mathcal{N}\) means that the next player to move has a winning strategy. Note that, since in a Maker-Breaker positional game, it is always better to go first rather than second, these are the only three possible outcomes. There is a direct connection between the outcomes of the scores:

\begin{remark}\label{dominator}
    Let \(G\) be a graph.
    \begin{itemize}
        \item The Domination Game on $G$ results in \(\mathcal{D}\) if and only if \(s(G) = 0\).
        \item The Domination Game on \(G\) results in \(\mathcal{N}\) if and only if \(Bs(G) = 0\) and \(Ms(G)\geq 1\).
        \item The Domination Game on \(G\) results in \(\mathcal{S}\) if and only if \(Bs(G) \ge 1\) and \(Ms(G)\geq 1\).
    \end{itemize}
\end{remark}

Note that this only makes it possible to differentiate $0$ scores from positive scores. For instance, if $G$ is a star with at least two leaves, the outcome of the Maker-Breaker domination game on $G$ is $\mathcal{N}$, but the score when Maker starts in the \textsc{SHVG} depends on the number of leaves.

In particular, if the outcome of a game is $\mathcal{D}$, it means that Breaker can ensure a score of $0$ going either first or second. Thus, using Theorem~\ref{adding game of fixed score}, we have the following corollary:

\begin{corollary}\label{neutral}
    Let \(G\) be a graph. Then, the outcome of the Domination game on \(G\) is \(\mathcal{D}\) if and only if \textsc{SHVG}\((G) = 0\).
\end{corollary}

A classical tool to prove that the outcome of the domination game is $\mathcal{D}$, is the \emph{pairing strategy}. This general notion exists in all Maker-Breaker positional games, and can be expressed as follows in the Maker-Breaker domination game: if $G = (V,E)$ is a graph, a set of disjoint pairs of vertices $(x_1, y_1), \dots, (x_k,y_k) \subset V^2$ is called a {\em pairing dominating set} if $V \subset \underset{1 \le i \le k}{\cup} N[x_i] \cap N[y_i]$. Duchêne {\em et al.}~\cite{duchene2020domination} proved that if $G$ has a pairing dominating set, then its outcome is $\mathcal{D}$. In our score-based framework, this leads to the following:

\begin{remark}\label{remark pds, scoring}
    Let $G$ be a graph. If $G$ has a pairing dominating set, then $G = 0$.
\end{remark}

In particular, this means that when considering disconnected graphs, removing a connected component that contains a pairing dominating set does not change the score on the game.

\subsection{Simplifying instances}

In this section, we aim to identify structures in a graph that can be reduced to other structures that are simpler to handle. In particular, when some vertices are already colored, we are looking for graph operations that will transform our graph into a simpler one without altering the score on the current position.

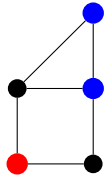
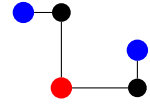
\begin{figure}[t!]
    \centering
    \begin{subfigure}[t]{0.49\textwidth}
        \centering

\begin{tikzpicture}

\draw (0,0) node[R] (a) {} node[below = .2]{};
\draw (1,0) node[v] (b) {} node[below = .2]{};
\draw (0,1) node[v] (c) {} node[below = .2]{};
\draw (1,1) node[B] (d) {} node[below = .2]{};
\draw (1,2) node[B] (e) {} node[below = .2]{};

\draw (a) -- (b);
\draw (d) -- (b);
\draw (d) -- (c);
\draw (a) -- (c);
\draw (e) -- (c);
\draw (d) -- (e);

\end{tikzpicture}
        
        \caption{A position $(G, M, B)$.}
    \end{subfigure}%
    ~ 
    \begin{subfigure}[t]{0.49\textwidth}
        \centering

\begin{tikzpicture}

\draw (0,0) node[R] (a) {} node[below = .2]{};
\draw (1,0) node[v] (b) {} node[below = .2]{};
\draw (0,1) node[v] (c) {} node[below = .2]{};
\draw (1,0.5) node[B] (d) {} node[below = .2]{};
\draw (-.5,1) node[B] (e) {} node[below = .2]{};

\draw (a) -- (b);
\draw (d) -- (b);
\draw (a) -- (c);
\draw (e) -- (c);

\end{tikzpicture}
        \centering

        \caption{The reduced position $([G]_B, [M]_B, [B]_B)$.}
    \end{subfigure}
    \caption{The reduction in Definition~\ref{decomposed graph}}.
    \label{fig: reduction decomposed graph}
\end{figure}

\begin{definition}\label{decomposed graph}
   Let \((G, M, B)\) define a position in the game on the graph \(G\). We define the \emph{decomposed graph} at this position, denoted by \([G]_B\), as the subgraph of \(G\) induced by the vertices \(M \cup V_F\), augmented as follows: for every vertex \(u\) in the subgraph induced by $M\cup V_F$ that is adjacent to at least one vertex \(v \in B\) in \(G\), we add a new vertex \(u_B\) and an edge \(\{u, u_B\}\). We denote by \([B]_B\) the set of all the vertices \(u_B\) added this way and \([M]_B=M\). This decomposition is depicted in Figure~\ref{fig: reduction decomposed graph}.
\end{definition}

This transformation consists in identifying the vertices dominated by Breaker, i.e. that cannot be made happy any longer, making them dominated only by a leaf, by removing all the remaining vertices claimed by Breaker.

\begin{lemma}\label{equiv decompose}
    Let \((G, M, B)\) define a position in the game on the graph \(G\). 
    
    Then, \(s(G, M, B) = s([G]_B, [M]_B, [B]_B)\).
\end{lemma}

\begin{proof}
Suppose Maker has a strategy that ensures a score $s_0$ in $G$. She can apply her strategy in \(([G]_B, [M]_B, [B]_B)\), which is possible since the vertices of $V_F([G]_B)$ are also free in $G$. At the end of the game, a vertex \(u\) is dominated in \(G\) if and only if it has a neighbor \(v\) colored by \Breaker{}. If this neighbor \(v\) is in \(B\), then \(u\) also has a neighbor colored by \Breaker{} in \([G]_B\). Otherwise, \(v\) belongs to \([G]_B\) and, as the strategy in $G[B]$ mimics the one in $G$, $v$ is still colored by \Breaker{} in $[G]_B$. Consequently, any happy vertex in \(G\) is also happy in \([G]_B\), ensuring that $s([G]_B) \ge s(G)$. Similarly, by applying his strategy in $G$ to $[G]_B$, Breaker can ensure that $s([G]_B) \le s(G)$, which concludes the proof.
\end{proof}

We now focus on properties that can help us computing the score, by limiting the number of moves to consider: When considering scoring game, it happens that some moves will always have a higher impact on the score than others. It then seems natural to think that these moves will be played first. Note that a similar Lemma is used in~\cite{bagan2024incidence}, but the structure of the winning sets being more complex in the Maker-Breaker happy vertex game, it's stated differently. We first introduce a function $h$ that takes a position $P = (G,M,B)$ and a free vertex $v \in V\setminus(M\cup B)$ and outputs the number of points scored instantly by playing $v$. Formally, we have: 

$$h(P,v) = \left|\{w \in G \mid N[w] \subseteq M \cup \{v\}\}\right|$$

If there is no ambiguity on $P$, we will simply denote it by $h(v)$. The next lemma says that if playing $v$ is worth more points than claiming $u$, even if all the neighbors of $u$ are made happy, then, there exists an optimal strategy in which $v$ is played before $u$.

\begin{lemma}\label{lemma:more-point-vertices}
    Let $(G,M,B)$ be a position of the game. Let $u, v$ be two vertices such that:

    $$ h(v) \ge |N[u] \setminus B| $$

    Then, there is an optimal strategy in which $v$ is played before $u$.
\end{lemma}

\begin{proof}
    We give the proof for Maker playing $u$ before $v$, a  similar proof provides the result if it is Breaker.

    Up to consider a position of the game that is reached later in the strategy, let $P = (G,M,B)$ be a position and suppose Maker has a strategy $\strat$ that scores $s_0$ in which the next move of Maker is to color $u$. Let $P' = (G,M\cup \{u\},B)$. We propose the following strategy for Maker:
\begin{itemize}
    \item Maker plays $v$.
    \item If Breaker colors a vertex $w \neq u$, Maker considers that he has played this vertex in $P'$. If the answer of Maker in $P'$ is a vertex $w' \ne v$, Maker answers by playing $w'$ in $P$. If Maker's answer in $P'$ is to color $v$, then Maker colors $u$ instead.
    \item If Breaker colors $u$, Maker considers that he has played $v$ in $P'$ and continue her strategy by playing move that she would have played according to $\strat$ (inverting the plays on $u$ and $v$).
\end{itemize}

Following this strategy, when all the vertices are colored, either Maker has colored both $u$ and $v$. In this case, the set of colored vertices by Maker and Breaker at the end of the game is exactly the same, so Maker scores $s_0$ by coloring $v$ first. Otherwise, the set of colored vertices is the same, except that Maker has colored $v$ instead of $u$, and Breaker has colored $u$ instead of $v$. Denote by $P_u = (G, M_u,B_u)$ and $P_v = (G, M_v, B_v)$ the resulting positions. The score in $P_v$ is then : 

\begin{align*}
    s(P_v) &= s_0 + |\{w \in N[v] | N[w] \subset M_v \} - |\{w \in N(u) | N[w] \subset M_u\}| \\
    &\ge s_0 + |\{w \in N[v] | N[w] \subset M \cup \{v\} \} - |\{w \in N(u) | N[u] \cap B =\emptyset\}|\\
    &\ge s_0
\end{align*}

Where the second inequality is obtained as we have $M \cup \{v\} \subset M_v$ and $M_u \subset V \setminus B$.

This proves that Maker can ensure a score at least $s_0$ by playing $v$ before $u$.
\end{proof}

The so-called \emph{Super Lemma} is a result that was introduced simultaneously in the PhD thesis of Nacim Oijid \cite{oijid2024thesis} and in the study of the game \textsc{Incidence} \cite{bagan2024incidence}. In this section, we adapt it to the \emph{Scoring Happy Vertex Game}, thereby extending its applicability and providing a crucial tool for analyzing the strategic structure of the game. In the context of our game, this lemma says that if two vertices have the same neighborhood, among the uncolored vertices, and scores the same number of points among the colored vertices, there exists an optimal strategy in which the two players \Maker{} and \Breaker{} will both color one of these two vertices. 

\begin{theorem}\label{super lemma}
Let \((G,M,B)\) be a position of the game on a graph \(G\), and let \(u,v \in V_F \) be two distinct vertices such that for any $X \subset V_F$, we have $$|\{w \in V| N[w] \subset X \cup \{u\} \cup M  \}| = |\{w | w \in V, N[w] \subset X \cup \{v\} \cup M  \}|$$
then
\[
s(G, M, B) = s(G, M \cup \{u\}, B \cup \{v\}).
\]
\end{theorem}

\begin{proof}

The proof is made by induction on $|V_F|$

If (\( \lvert V_F \rvert = 2 \)), we have \( V_F = \{u, v\} \). For \( X = \emptyset \), the hypothesis implies:
$$|\{w \in V| N[w] \subset \{u\} \cup M  \}| = |\{w | w \in V, N[w] \subset \{v\} \cup M  \}|$$
Therefore, after \Maker{} and \Breaker{}'s turns, the score is the same regardless of the choice of the vertex colored by the next player. Hence,
\[
s(G, M, B) = s(G, M \cup \{u\}, B \cup \{v\}).
\]

Suppose now $|V_F| \ge 3$. Assume first that \Maker{} has a strategy $\strat$ to ensure a score $s_0$ in $(G, M \cup \{u\}, B \cup \{v\})$, and consider the following strategy in $(G,M,B)$:

\begin{itemize}
    \item If it's Maker's turn, she colors the vertex she would have color in $(G, M \cup \{u\}, B \cup \{v\})$, following $\strat$.
    \item If Breaker colors a vertex in $\{u,v\}$, Maker colors the second vertex of $\{u,v\}$, which leads to the same score as in $G'$ by hypothesis on $\{u,v\}$.
    \item Otherwise, If Breaker colors $w \notin \{u,v\}$ Maker colors the vertex $w'$ that she is supposed to color according to $\strat$, considering that the last move of Breaker in  $(G, M \cup \{u\}, B \cup \{v\})$ was $w$. Note that this move is always available and cannot be a vertex \{u,v\}, unless $|V_F| = 3$, in which case, Breaker would also color $w$ in $(G, M \cup \{u\}, B \cup \{v\})$.
\end{itemize}

Following this strategy, if the result is not obtained directly, the resulting position is $( G, M\cup\{w'\}, B \cup \{w\})$ which has two less free vertices. Hence, by induction, it has the same score as $(G, M \cup \{u,w'\}, B \cup \{v,w\})$, and since the moves $w'$ was supposed to be optimal answer to $w$ in $(G, M \cup \{u\}, B \cup \{v\})$, it has a score equal or higher than $(G, M \cup \{u\}, B \cup \{v\})$, which proves that Maker can ensure $s_0$ in $(G,M,B)$.

A similar argument applied to Breaker shows that if Breaker can ensure a score of at most $s_0$ in $(G, M \cup \{u\}, B \cup \{v\})$, he can ensure at most $s_0$ in $(G,M,B)$, which concludes the proof.
\end{proof}

Similarly to the provided result in~\cite{bagan2024incidence}, this lemma enables us to directly compute the score on {\em complete binary trees}, i.e. trees defined by $T_0$ a single vertex, which is a root, and $T_i$ is obtained by taking two copies of $T_{i-1}$, and adding a new root that will be connected to the previous roots of the copies of $T_{i-1}$. The integer $i$ is then called the {\em depth} of $T_i$.

\begin{theorem}
    Let $T_d$ be the complete binary tree of depth $d$. We have $Ms(T_d) = 1$ and $Bs(T_d) = 0$ if $d \le  1$, and $Ms(T_d) = Bs(T_d) = 2^{d-2}$ otherwise. 
\end{theorem}

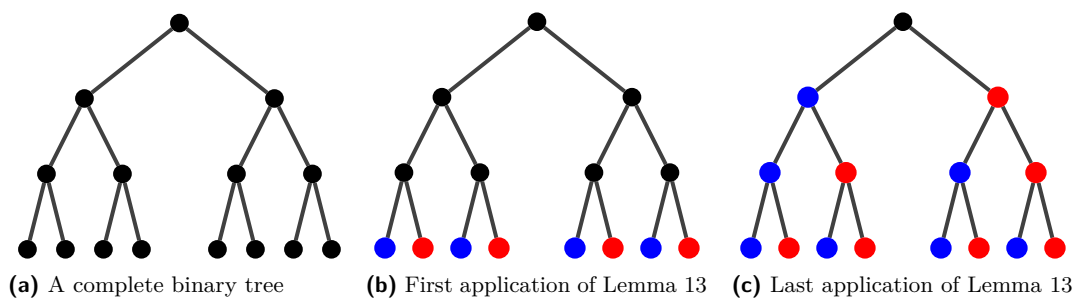
\begin{figure}[ht!]
    \centering
    \begin{subfigure}[t]{0.32\textwidth}
        \centering

\begin{tikzpicture}
\draw (2,4) node[v](10){} ;

\draw (3.25,3) node[v](20){} ;
\draw (.75,3) node[v](21){} ;

\draw (3.75,2) node[v](30){} ;
\draw (2.75,2) node[v](31){} ;
\draw (1.25,2) node[v](32){} ;
\draw (0.25,2) node[v](33){} ;

\draw (4,1) node[v](40){} ;
\draw (3.5,1) node[v](41){} ;
\draw (3,1) node[v](42){} ;
\draw (2.5,1) node[v](43){} ;
\draw (1.5,1) node[v](44){} ;
\draw (1,1) node[v](45){} ;
\draw (.5,1) node[v](46){} ;
\draw (0,1) node[v](47){} ;

\Edge[](10)(20)
\Edge[](10)(21)

\Edge[](20)(30)
\Edge[](20)(31)
\Edge[](21)(32)
\Edge[](21)(33)

\Edge[](30)(40)
\Edge[](30)(41)
\Edge[](31)(42)
\Edge[](31)(43)
\Edge[](32)(44)
\Edge[](32)(45)
\Edge[](33)(46)
\Edge[](33)(47)

\end{tikzpicture}
        
        \caption{A complete binary tree }
    \end{subfigure}%
    ~ 
    \begin{subfigure}[t]{0.32\textwidth}
        \centering

\begin{tikzpicture}
\draw (2,4) node[v](10){} ;

\draw (3.25,3) node[v](20){} ;
\draw (.75,3) node[v](21){} ;

\draw (3.75,2) node[v](30){} ;
\draw (2.75,2) node[v](31){} ;
\draw (1.25,2) node[v](32){} ;
\draw (0.25,2) node[v](33){} ;

\draw (4,1) node[R](40){} ;
\draw (3.5,1) node[B](41){} ;
\draw (3,1) node[R](42){} ;
\draw (2.5,1) node[B](43){} ;
\draw (1.5,1) node[R](44){} ;
\draw (1,1) node[B](45){} ;
\draw (.5,1) node[R](46){} ;
\draw (0,1) node[B](47){} ;

\Edge[](10)(20)
\Edge[](10)(21)

\Edge[](20)(30)
\Edge[](20)(31)
\Edge[](21)(32)
\Edge[](21)(33)

\Edge[](30)(40)
\Edge[](30)(41)
\Edge[](31)(42)
\Edge[](31)(43)
\Edge[](32)(44)
\Edge[](32)(45)
\Edge[](33)(46)
\Edge[](33)(47)

\end{tikzpicture}

        \caption{First application of Lemma~\ref{super lemma}}
    \end{subfigure}
    ~
    \begin{subfigure}[t]{0.32\textwidth}
        \centering

\begin{tikzpicture}
\draw (2,4) node[v](10){} ;

\draw (3.25,3) node[R](20){} ;
\draw (.75,3) node[B](21){} ;

\draw (3.75,2) node[R](30){} ;
\draw (2.75,2) node[B](31){} ;
\draw (1.25,2) node[R](32){} ;
\draw (0.25,2) node[B](33){} ;

\draw (4,1) node[R](40){} ;
\draw (3.5,1) node[B](41){} ;
\draw (3,1) node[R](42){} ;
\draw (2.5,1) node[B](43){} ;
\draw (1.5,1) node[R](44){} ;
\draw (1,1) node[B](45){} ;
\draw (.5,1) node[R](46){} ;
\draw (0,1) node[B](47){} ;

\Edge[](10)(20)
\Edge[](10)(21)

\Edge[](20)(30)
\Edge[](20)(31)
\Edge[](21)(32)
\Edge[](21)(33)

\Edge[](30)(40)
\Edge[](30)(41)
\Edge[](31)(42)
\Edge[](31)(43)
\Edge[](32)(44)
\Edge[](32)(45)
\Edge[](33)(46)
\Edge[](33)(47)

\end{tikzpicture}

        \caption{Last application of Lemma~\ref{super lemma}}
    \end{subfigure}
    \caption{Computing the score on $T_3$, the complete binary tree of depth $3$.}
    \label{fig: complete binary tree}
\end{figure}

\begin{proof}
    If $d = 0$, $T_d$ is an isolated vertex, and Maker scores one by coloring it, otherwise the score is $0$. If $d = 1$, the graph is $P_3$, the path on $3$ vertices. We can apply Lemma~\ref{super lemma} to its two leaves, and Maker scores $1$ if she starts by coloring the middle vertex; otherwise, Breaker colors it and ensures that no vertex is happy. If $d \ge 2$, each pair of leaves attached to the same vertex are twins, thus, we can apply Lemma~\ref{super lemma} to them. Hence, Maker has colored half of them, i.e. $2^{d-1}$ of them. Now their parents satisfy the hypothesis of Lemma~\ref{super lemma}, as they are in the winning set of the connected leaf colored by Maker and the winning set of their common parent (which exists as $d \ge 2$). Thus, Maker will color half of them, ensuring that half of the leaves she has colored are happy, and thus scoring $2^{d-2}$. We can then repeat this process until reaching the vertices of depth $1$, but now, all the colored vertices by Maker already have one neighbor colored by Breaker, and thus will not be happy. Finally, regardless of whether Maker or Breaker colors the root, it has one neighbor colored by Breaker and thus will not be happy. Finally, the score is $2^{d-2}$. This process is depicted in Figure~\ref{fig: complete binary tree}.
\end{proof}

\section{Hardness results}\label{sec: hardness}


In this section, we concern ourselves with the hardness of the \textsc{SHVG} problem. Note that, since the Maker-Breaker domination game is already \PSPACE-complete on bounded degree bipartite graphs, or split graphs~\cite{duchene2020domination, oijid2025qbf}, we can inherit this result for $s = 1$. Namely, we have the following theorem:

\begin{theorem}
    \textsc{SHVG} is \PSPACE-complete even restricted to $s = 1$, and to bounded degree bipartite graphs or split graphs.
\end{theorem}

However, adding scores to the Maker-Breaker domination game makes it hard on smaller graph classes, and we prove that \textsc{SHVG} is already \PSPACE-complete when restricted to trees, and that the problem is \NP-hard when restricted to caterpillars.

Both hardness results will rely on reductions from two specific Quantified MAX-SAT problems, which we introduce and prove to be \PSPACE-complete and at least \NP-hard respectively in the next subsection. We postpone the hardness results on \textsc{SHVG} to Section~\ref{sec:hardness-trees} and Section~\ref{sec:hardness-caterpillars}. 

\subsection{Quantified MAX SAT restrictions} \label{sec:quantified-max-sat}

In this subsection, we will introduce two restrictions of the \textsc{(Quantified) MAX-$2$-SAT} problem which we will use in order to obtain our hardness results. These problems are very general, and an important contribution of this paper since they can be used to provide other reduction, in particular, providing a natural \PSPACE-complete problem on forests.
Recall first the definition of \textsc{Quantified MAX 2-SAT}, which is a quantified version of \textsc{Max $2$-SAT}.

\begin{problem}[\textsc{Quantified Max 2-SAT}]\label{qmaxsat}{}
\hspace{.2cm}

\noindent    
\Input: $(\psi, k)$ where $\psi$ is a QBF formula of the form $\psi = Q_1 x_1 Q_2 x_2 \ldots Q_{n}x_{n} ~ \varphi$ with, for $1 \le i \le n$, $Q_i \in \{\exists, \forall\}$, $\varphi$ a quantifier free $2$-CNF formula, and $k$ is an integer.
    
\noindent    
\Question: 
Is it possible to satisfy simultaneously $k$ clauses of $\varphi$?
\end{problem}

As usual dealing with positional games, we will consider the equivalent gaming variant of the problem~\cite{stockmeyer1973, bagan2024incidence}: Two players Satisfier and Falsifier choose the value of the $x_i$s, Satisfier chooses the existentially quantified variables, and Falsifier chooses the universally quantified variables. Satisfier wins if she manages to satisfy $k$ clauses, otherwise Falsifier wins.


\textsc{Quantified MAX 2-SAT} is \PSPACE-complete~\cite{bagan2024incidence}. 
We will prove that this remains the case even when we impose structural properties on the formula $\psi$. 

\begin{figure}
   \begin{center}
\begin{tikzpicture}[
          var/.style={circle,draw=gray,thick,fill=gray!20,minimum size=8mm},
          clause/.style={rectangle,draw=gray,thick,fill=gray!10,minimum size=8mm},
          negclause/.style={rectangle,draw=gray,thick,dashed,fill=gray!10,minimum size=8mm},
          >=latex
        ]
        \node[clause] (X1+) at (0,0) {$x_1^+$};
        \node[negclause] (X1-) at (0,-2) {$x_1^-$};
        \node[clause] (X2+) at (3,0) {$x_2^+$};
        \node[negclause] (X2-) at (3,-2) {$x_2^-$};
        \node[clause] (X3+) at (6,0) {$x_3^+$};
        \node[negclause] (X3-) at (6,-2) {$x_3^-$};
        
        \draw (X1+) -- (X2-);
        \draw (X2+) -- (X3+);
        \draw (X1-) -- (X3+);
        \end{tikzpicture}
    \end{center}
    \caption{The literal-clause incidence graph $G_\varphi$ with \(\varphi=(x_1\lor\lnot x_2)\land(x_2\lor x_3)\land(\lnot x_1\lor x_3)\)}
    \label{fig:literal-clause incidence graph}
\end{figure}
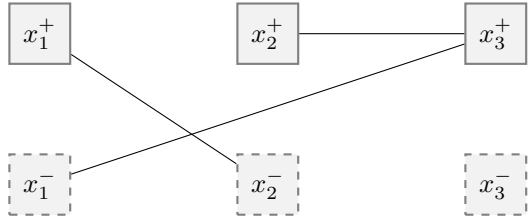

We now define the main tool that will be used in our reductions. Let $\varphi$ be a quantifier-free $2$-CNF formula, with variables \(\{x_1, \dots, x_n\}\). Let \( G_{\varphi}\) be the graph in which each variable \(x_i\) is represented by two vertices, denoted \(x_i^+\) and \(x_i^-\), corresponding to its positive and negative literals respectively, and where for each clause \(l_i \lor l_j\) in \(\varphi\), there is an edge between the vertices corresponding to the literals \(l_i\) and \(l_j\).
We will call $G_{\varphi}$ the {\em literal–clause incidence graph} of $\varphi$.  An example of construction of this graph is provided in Figure~\ref{fig:literal-clause incidence graph}. For readers used to expressing boolean formulas in graphs, note that this graph is {\bf not} the usual incidence graph of a formula.

Let us now introduce our first specific Quantified MAX-$2$-SAT problem.

\begin{samepage}
    \begin{problem}[\textsc{Acyclic Q-MAX 2-SAT}]\label{qmax2sat}{}
\hspace{.2cm} \\
\noindent    
\Input: A QBF formula of the form $\psi = Q_1 x_1 Q_2 x_2 \ldots Q_{n}x_{n} ~ \varphi$ with $\varphi$ a quantifier-free $2$-CNF formula whose literal–clause incidence graph is acyclic, an integer $k$. \\
\noindent    
\Question: 
Is it possible to satisfy simultaneously $k$ clauses of $\psi$?
\end{problem}
\end{samepage}

\begin{lemma}
    \textsc{Acyclic Q-MAX 2-SAT} is \(\mathsf{PSPACE}\)-complete.
\end{lemma}
\begin{proof}

Membership in \PSPACE~follows from the  \PSPACE-completeness of \textsc{Quantified MAX 2-SAT}~\cite{bagan2024incidence}. To prove the hardness, we reduce from \textsc{Quantified MAX 2-SAT} without the acyclicity restriction. Let $(\psi, k)$ be an instance of  \textsc{Quantified MAX 2-SAT}, with  $\psi = Q{x_1} Q{x_2} \ldots Q{x_n} ~ \varphi$ where $\varphi$ is a quantifier-free formula.
Let $(l_1 \lor l_2) \land (l_2 \lor l_3) \land \dots \land (l_m \lor l_1)$ be a cycle in $\varphi$. Without loss of generality, assume that \(l_1 = x\), the case $l_1 = \neg x$ being similar. We then introduce three new existential quantifiers \(\exists a \exists b \exists x'\) and the following gadget clauses:  
\[
(x \lor a), \quad (\lnot x \lor b), \quad (\lnot a \lor \lnot x'), \quad (\lnot b \lor x').
\]

We consider the formula $\psi' = Q{x_1} Q{x_2} \ldots Q{x_n} ~\exists x' \exists a \exists b  ~ \varphi'$ where $\varphi'$ is obtained from $\varphi$ by substituting $(x' \lor l_2)$ to $(x \lor l_2)$, and by adding the clauses 
$(x \lor a),  (\lnot x \lor b),  (\lnot a \lor \lnot x'),  (\lnot b \lor x')$. Let $k' = k +4$.

The intuition behind this construction is that if there is an assignment of the variables that satisfy $k$ clauses in $ \psi$, the same assignment putting $x' = x$, and choosing the value of $a$ and $b$ such that $a = b = \neg x$ will satisfy $k+4$ clauses in $\psi'$.

Suppose first that Satisfier has a winning strategy in $(\psi,k)$, by applying the same strategy in $(\psi',k')$, and by setting $x' = x$, she ensures that at least $k+2$ clauses are satisfied when $x'$ is played, as exactly one of $(x\vee a), (\neg a \vee \neg x')$ and exactly one of $(\neg x\vee b), (\neg b \vee x')$. Then, by putting $a = b = \neg x$, she ensures to satisfy the two remaining clauses, and that at least $k+4$ clauses are satisfied.

Reciprocally, suppose that Falsifier has a winning strategy in $(\psi, k)$. By applying the same strategy in $(\psi', k')$, he ensures that at most $k-1$ clauses are satisfied when $x_n$ is played. Then, if Satisfier puts $x = x'$, the only clauses that can be satisfied in $\psi'$ that are not in $\psi$ are the $4$ added clauses, hence, he ensures that at most $k+3 < k'$ are satisfied in total. If Satisfier sets $x' = \neg x$, the only clause that can be satisfied that was in $\psi$ through this move is the clause $x' \vee l_2$, since the other clauses do not contain $x'$. Hence, at most $k$ clauses are satisfied outside  $\{(x \lor a), (\lnot x \lor b), (\lnot a \lor \lnot x'), (\lnot b \lor x')\}$. But, either $x$ and $\neg x'$ are True, then, at most one of  $\{(\lnot x \lor b), (\lnot b \lor x')\}$ can be satisfied, or $x$ and $\neg x'$ are False, and then at most one of $\{ (x \lor a),
(\lnot a \lor \lnot x')\}$ can be satisfied. In both cases, at most three of these clauses can be satisfied, and thus, at most $k+3 < k'$ in total, ensuring that Falsifier wins in $(\psi',k')$. Hence, Satisfier wins in $(\psi, k)$ if and only if she wins in $(\psi',k')$

Finally, in the constructed formula $\psi'$, the literal–clause incidence graph has at least one less cycle than the one of $\psi$ since the variables \(x\) and \(x'\) are separated by the variables \(a\), \(\lnot a\) and \(b\), \(\lnot b\).

By iterating this construction we can eliminate all cycles. Note that each time this reduction is applied, the number of clauses that can be in a cycle decreases by one, since none of the added clauses can be in a cycle. Thus, the number of time this argument has to be applied is bounded by the number of clause of $\psi$, ensuring that this reduction is polynomial.
Therefore, \textsc{Acyclic Q-MAX 2-SAT} is \PSPACE-complete.
\end{proof}

Then, we can define our second more specific problem, where we restrict the number of times a variable can occur in a clause.

\begin{samepage}
\begin{problem}[\textsc{Acyclic MAX 2-SAT-2-2}]\label{max2sat22}{}
\hspace{.2cm}

\noindent    
\Input: A $2$-CNF formula $\varphi$ such that
\begin{itemize}
    \item the literal–clause incidence graph $G_\varphi$ is acyclic,
    \item each variable occurs at most twice positively and at most twice negatively,
\end{itemize} an integer $k$.
    
\noindent    
\Question: 
Is it possible to satisfy simultaneously $k$ clauses of $\varphi$?
\end{problem}
\end{samepage}

This problem consists in a quantified version of bounded occurrence 
\textsc{MAX 2-SAT} - but where we furthermore impose acyclicity on the literal-clause incidence graph.
First, we recall that \textsc{MAX 2-SAT-3}, the version of \textsc{MAX 2-SAT} where the number of occurrences of each variable is bounded by $3$ is \NP-complete~\cite{berman2002some}.    
More formally, this means that the following problem is \NP-complete:

\begin{samepage}
\begin{problem}[\textsc{MAX 2-SAT-3}]\label{max2sat3}{}
\hspace{.2cm}

\noindent    
\Input: A $2$-CNF formula $\varphi$ such that each variable occurs at most thrice,
 an integer $k$.
    
\noindent    
\Question: 
Is it possible to satisfy simultaneously $k$ clauses of $\varphi$?
\end{problem}
\end{samepage}

We now prove the \NP-completeness of \textsc{Acyclic MAX 2-SAT-2-2} by a reduction from \textsc{MAX 2-SAT-3}.

\begin{lemma}\label{MAX 2-SAT-2-2 acyclic}
   \textsc{Acyclic MAX 2-SAT-2-2} is \NP-complete. 
\end{lemma}
\begin{proof}
First, note that  \textsc{Acyclic MAX 2-SAT-2-2} is in \NP\ as it is an instance of \textsc{Max 2-SAT}.

We reduce from the \NP-complete problem \textsc{MAX 2-SAT-3}
~\cite{berman2002some}. Let $\varphi$ be a $2$-CNF formula such that each variable occurs at most thrice and $k$ an integer.
If a variable in $\varphi$ appears more than twice positively (resp. negatively) then it must appears exactly thrice positively (resp. negatively). Then we can satisfy all three clauses by assigning the variable to \texttt{True} (resp. \texttt{False}) and no clause is satisfied by changing the truth value of the variable from \texttt{True} to \texttt{False} (resp. False to \texttt{True}), i.e. we can always satisfy these three clauses without incidence on the other clauses. Thus we can simply remove these three clauses from $\varphi$ and lower $k$ by $3$. By iterating this process we construct a new formula where variables not only occur at most thrice, but also occur at most twice positively and at most twice negatively.

Furthermore, if a variable appears in a cycle in the implication graph, we can apply the gadget from the previous reduction to effectively remove cycles. However, this gadget introduces a new variable \(x_i'\) for each \(x_i\), and requires using one more occurrence of \(x_i\) if $x_i$ appears negatively in the cycle or one more of \(\lnot x_i\) if it appears positively in the cycle. However, this transformation cannot make a variable appear a third time positively or negatively, since if $x_i$ ($\neg x_i$ resp.) appears in the cycle, it means that it appears exactly twice positively (negatively resp.) and thus at most once negatively (positively resp.). Note also that this transformation ensures that all the occurences of $x_i$ and $\neg x_i$ are either of degree~$1$, or neighbors of a vertex of degree~$1$, ensuring that they cannot appear in cycle, and thus this reduction has to be applied at most once per variable. Hence, this transformation does not make any variable appear more than twice positively or negatively. This entire transformation can be performed in polynomial time, therefore we conclude that the problem is \NP-complete.
\end{proof}

A direct consequence of this result is that, by quantifying all the vertices with $\exists$, it's quantified version is \NP-hard and in \PSPACE.

\begin{samepage}
\begin{problem}[\textsc{Acyclic QMAX 2-SAT-2-2}]\label{qmax2sat22}{}
\hspace{.2cm}

\noindent    
\Input: A quantified $2$-CNF formula $\psi = Q_1 x_1\dots Q_n x_n \varphi$ such that
\begin{itemize}
    \item the variable–clause incidence graph $G_\varphi$ is acyclic,
    \item each variable occurs at most twice positively and at most twice negatively,
\end{itemize} an integer $k$.
    
\noindent    
\Question: Can Satisfier satisfy simultaneously $k$ clauses of $\psi$?
\end{problem}
\end{samepage}

\begin{corollary}\label{QMAX 2-SAT-2-2 acyclic}
       \textsc{Acyclic QMAX 2-SAT-2-2} is \NP-hard and in \PSPACE.
\end{corollary}

\begin{proof}
    The \NP-hardness is directly inherited from \textsc{Acyclic MAX 2-SAT-2-2} by quantifying all the variables with $\exists$. The problem is in \PSPACE\ as a subproblem of \textsc{QMAX 2-SAT}.
\end{proof}

These new variants of \textsc{Q-MAX 2-SAT} are quite general and can be used in several reductions. In particular, the reduction from \textsc{Q-MAX 2-SAT} provided in~\cite{bagan2024incidence}, proving that Incidence, the scoring positional game on graphs is \PSPACE-complete starts from the literal-clause incidence graph of a formula and only adds isolated vertices and leaves to it. Hence, we almost answer an open problem of the paper, that asks the complexity of the game on trees with the following corollary:

\begin{corollary}\label{corollary Incidence pspace complete forests}
    \textsc{Maker-Breaker Incidence} is \PSPACE-complete, even restricted to forests, and \NP-hard, even restricted to unions of caterpillars.
\end{corollary}

\begin{proof}
    We perform the same reduction as in~\cite{bagan2024incidence}, except that the input problem is \textsc{Acyclic QMAX 2-SAT} (or \textsc{Acyclic QMAX 2-SAT-2-2}). Since this constructions consists in adding leaves to the literal-clause incidence graph of the formula, it produces either a forest or a caterpillar, depending on the problem from which we do the reduction.
\end{proof}

\subsection{Trees}\label{sec:hardness-trees}

We now have all the necessary tools to establish the \PSPACE-completeness of \textsc{SHVG} when restricted to trees. 
The proof relies on a reduction from \textsc{Acyclic Q-MAX 2-SAT} which we proved to be \PSPACE-hard.
Since \textsc{SHVG} belongs to \PSPACE~ as it is a scoring positional game~\cite{bagan2024incidence}, the result follows.

\begin{theorem}\label{complexity trees}
    \textsc{SHVG} on trees is \PSPACE-complete.
\end{theorem}
\begin{proof}
We provide a reduction from \textsc{Acyclic Q-MAX 2-SAT}. Let $(\psi,k)$ be an instance of \textsc{Acyclic Q-MAX 2-SAT} with $\psi = Q_1{x_1} Q_2{x_2} \ldots Q_n{x_n} ~ \varphi$ with $\varphi$ an acyclic and quantifier-free $2$-CNF formula. Up to add variables that appear in no clauses, and thus that do not break the acyclicity of the literal-clause incidence graph, we can suppose that $Q_i = \exists$ if $i$ is odd, and $Q_i = \forall$ if $i$ is even.
The first step of our reduction is to duplicate our formula: we transform it into $\psi' = Q_1{x_1} Q_2{x_2} \ldots Q_n{x_n} Q_{n+1}{x_{n+1}} Q_{n+2}{x_{n+2}} \dots Q_{2n}{x_{2n}} ~ \varphi \wedge \varphi'$, where $\varphi'$ is obtained from $\varphi$ by changing each $x_i$ to $x_{n+i}$. Note that any strategy satisfying $p$ clauses can be applied twice to satisfy $2p$ clauses in $\psi'$. Let $m$ be the number of clauses of $\varphi \wedge \varphi'$ We will progressively construct a tree by making modifications to the literal-clause incidence graph $G_{\varphi \wedge \varphi'}$ of $\varphi \wedge \varphi'$.

Since the graph is acyclic by hypothesis, it must be a disjoint union of trees $T_1, T_2, \ldots T_\ell$. We now construct a tree $T$ based on this graph. We start from $T_1 \cup T_2 \cup \dots \cup T_\ell$.
For every tree $T_i$ let us fix two of its leaves $v_i$ and $u_i$ and let us connect all trees into a single one by adding edges between $u_i$ and $v_{i+1}$ for $1 \leq i < \ell$. 
We denote by $\sim$ either $+$ or $-$. For each edge \(\{v_i^\sim, v_j^\sim\}\), we introduce a new vertex \(v_{ij}^{\sim\sim}\) and replace the edge with two edges: \(\{v_i^\sim, v_{ij}^{\sim\sim}\}\) and \(\{v_{ij}^{\sim\sim}, v_j^\sim\}\). 

Next, to each vertex \(v_i^\sim\), we attach \(16(n+1-i) m\) leaves.  Finally, we introduce \(n\) new vertices \(v_i\),  attach \(16(n+1-i)m-4m\) leaves to each of them, add edges between them, and between $v_n$  and $v_1$. 

An example of reduction is provided in Figure~\ref{fig:tree reduction}.

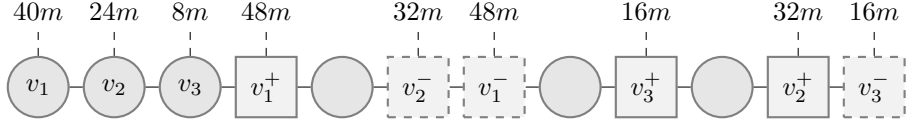
\begin{figure}
   \begin{center}
\begin{tikzpicture}[
scale=0.5,
          var/.style={circle,draw=gray,thick,fill=gray!20,minimum size=8mm},
          clause/.style={rectangle,draw=gray,thick,fill=gray!10,minimum size=8mm},
          negclause/.style={rectangle,draw=gray,thick,dashed,fill=gray!10,minimum size=8mm},
          >=latex
        ]
        \node[clause] (X1+) at (6,0) {$v_1^+$};
        \node[negclause] (X1-) at (12,0) {$v_1^-$};
        \node[clause] (X2+) at (20,0) {$v_2^+$};
        \node[negclause] (X2-) at (10,0) {$v_2^-$};
        \node[clause] (X3+) at (16,0) {$v_3^+$};
        \node[negclause] (X3-) at (22,0) {$v_3^-$};
        \node[var] (X12+-) at (8,0) {};
        \node[var] (X23++) at (18,0) {};
        \node[var] (X13-+) at (14,0) {};
        \node[var] (X1) at (0,0) {$v_1$};
        \node[var] (X2) at (2,0) {$v_2$};
        \node[var] (X3) at (4,0) {$v_3$};

        \node (k1) at (0,2) {$40m$};
        \node (k2) at (2,2) {$24m$};
        \node (k3) at (4,2) {$8m$};
        \node (k1+) at (6,2) {$48m$};
        \node (k2-) at (10,2) {$32m$};
        \node (k1-) at (12,2) {$48m$};
        \node (k3+) at (16,2) {$16m$};
        \node (k2+) at (20,2) {$32m$};
        \node (k3-) at (22,2) {$16m$};

        \draw[dashed] (X1) -- (k1);
        \draw[dashed] (X2) -- (k2);
        \draw[dashed] (X3) -- (k3);
        \draw[dashed] (X1+) -- (k1+);
        \draw[dashed] (X2-) -- (k2-);
        \draw[dashed] (X1-) -- (k1-);
        \draw[dashed] (X3+) -- (k3+);
        \draw[dashed] (X2+) -- (k2+);
        \draw[dashed] (X3-) -- (k3-);
        
        \draw (X1) -- (X2);
        \draw (X2) -- (X3);
        \draw (X1+) -- (X12+-);
        \draw (X12+-) -- (X2-);
        \draw (X2+) -- (X23++);
        \draw (X23++) -- (X3+);
        \draw (X1-) -- (X13-+);
        \draw (X13-+) -- (X3+);
        \draw (X3) -- (X1+);
        \draw (X2-) -- (X1-);
        \draw (X2+) -- (X3-);
        \end{tikzpicture}
    \end{center}    
    \caption{The resulting tree obtained from the formula $\exists x_1 \forall x_2 \exists x_3 ~ (x_1 \vee \neg x_2) \wedge (\neg x_1 \vee x_3) \wedge (x_3 \vee x_2)$}
    \label{fig:tree reduction}
\end{figure}

Let $s_0$ be defined as follows:

$$ s_0 = \left (\underset{1 \le i \le 2n}{\sum} 8(n+1-i) m \right) + \left ( \underset{1 \le i \le n}{\sum} 8\big (n+1-(2i-1)\big)m - 4m \right )$$

Let $s = s_0 + m - k +1$. Both $T$ and $s$ can be constructed in polynomial time. We prove that Falsifier wins in $(\psi,k)$ if and only if Maker wins in $(T,s)$.

Firstly, each leaf can be colored using Lemma~\ref{super lemma}. Denote by $M$ and $B$ the vertices colored during this step. Then, if $1 \le i \le j \le 2n$, we have $h(v_i^\sim) \ge |N(v_i) \setminus B|$ and $ h(v_i)  \ge |N(v_j^\sim) \setminus B|$. Hence, using Lemma~\ref{lemma:more-point-vertices}, there exists an optimal strategy in which \Maker{} and \Breaker{} color the vertices in the following order, for $1 \le i \le n$, in increasing order: 
\begin{itemize}
    \item Maker colors one of $\{v_{2i-1}^+, v_{2i-1}^-\}$.
    \item Breaker colors the other vertex in $\{v_{2i-1}^+, v_{2i-1}^-\}$.
    \item Maker colors $v_{2i-1}$.
    \item Breaker colors one of $\{v_{2i}^+, v_{2i}^-\}$.
    \item Maker colors the other vertex in $\{v_{2i}^+, v_{2i}^-\}$.
    \item Breaker colors $v_{2i}$.
\end{itemize}

Considering this order, which is optimal, note that all the moves are forced or have to be chosen among two vertices $(v_i^+, v_i^-)$ that are connected to the same number of leaves. Moreover, no vertex $v_i^\sim$ nor $v_i$ can be happy, because of the leaves colored by Breaker connected to them. Hence, we know that after this step, Maker has colored exactly $s_0$ happy vertices.

Finally, once all the previously described vertices have been colored, only the vertices \(v_{ij}^{\sim\sim}\) remain. These vertices will be colored last, and each of them can be happy if and only if both of their adjacent vertices have been colored by \Maker{}. 

Suppose first that Falsifier has a winning strategy $\strat$ in $(\psi,k)$. We define the following strategy for Maker:
\begin{itemize}
    \item For $0 \le i \le n-1$, if Falsifier has to put $x_{2i+1}$ to True, Maker colors $v_i^-$, ensuring that Breaker will color $v_i^+$; otherwise, she colors $v_i^+$, ensuring that Breaker colors $v_i^-$. 
    \item For $1 \le i \le n$, if Breaker colors $v^+_{2i}$, Maker considers that Satisfier has put $x_{2i}$ to True in $\strat$. Otherwise, she considers that Satisfier has put $x_{2i}$ to False.
    \item When $x_n$ is played, Maker applies $\strat$ again from scratch.
\end{itemize}

When all the vertices $v_i, v_i^+$ and $v_i^-$ have been played, if $C_j$ is a clause that has not been satisfied in $\varphi \wedge \varphi'$, then its two literals has been put to False. Thus, since $\strat$ was a winning strategy for Satisfier in $(\psi,k)$, we know that at most $k-1$ clauses of $\varphi$ and at most $k-1$ clauses of $\varphi'$ are satisfied. Thus, at least $2 \big( m-(k-1) \big)$ vertices $v_{ij}^{\sim\sim}$ has their two neighbors colored by Maker. Hence, Maker can by playing only on them ensure to color at least half of them, i.e. $m-k+1$, and having colored $s_0 + m - k + 1 = s$ happy vertices in total.

Reciprocally, suppose that Satisfier has a winning strategy $\strat$ in $(\psi,k)$. We define the following strategy for Breaker:
\begin{itemize}
    \item For $0 \le i \le n-1$, if Satisfier has to put $x_{2i+1}$ to True, Breaker colors $v_i^+$; otherwise, he colors $v_i^-$. 
    \item For $1 \le i \le n$, if Maker colors $v^+_{2i}$, Breaker colors $v^-{2i}$ and considers that Satisfier has put $x_{2i}$ to False in $\strat$. Otherwise, she considers that Satisfier has put $x_{2i}$ to True.
    \item When $x_n$ is played, Breaker applies $\strat$ again from scratch.
\end{itemize}

Now, when all the vertices,  $v_i, v_i^+$ and $v_i^-$ have been played, the same argument as above ensures that any clause that was satisfied by $\strat$ has one of its literal played by Breaker. Hence, since $\strat$ was a winning strategy in $\psi$, at least $2k$ vertices $v^{\sim\sim}_{ij}$ are already dominated by Breaker, and thus, at most $2(m-k)$ of them are not. By pairing them, Breaker can ensure to claim at least $m-k$ of them ensuring that Maker scores at most $s_0 + m -k < s$. Thus, Breaker wins.
\end{proof}

\subsection{Caterpillars}\label{sec:hardness-caterpillars}

We now consider \textsc{SHVG} restricted to caterpillars. A graph is called a \emph{caterpillar} if removing all its leaves results in a path. In other words, a caterpillar consists of a central path with pendant vertices attached to some or all of its nodes. 

\begin{theorem}\label{caterpillars complexity}
\textsc{SHVG} on caterpillars is \(\mathsf{NP}\)-hard.
\end{theorem}

\begin{proof}
    The proof is identical to the one on trees, except that instead of using \textsc{Acyclic Q-MAX 2-SAT}, we now rely on \textsc{Acyclic Q-MAX 2-SAT-2-2}, which is proved to be \NP-hard. In this case, the constructed graph  \( G_{\varphi}\) is not only a disjoint union of trees, but in particular a disjoint union of paths, since each vertex in the graph would have degree at most \(2\). The same reduction carries through and produces 
    a caterpillar. This shows that \textsc{SHVG} remains \NP-hard on caterpillars.
\end{proof}

\section{Efficient algorithms}\label{sec: poly}

Since the game is already \PSPACE-complete on forests, and \NP-hard on caterpillars, we aim to find efficient algorithms on other subclasses of trees. Namely, we focus here on union of paths and subdivided stars.

\subsection{Union of paths}

We first consider union of paths.

\subsubsection{Single path}

We begin by studying paths. A path $P_n$ is defined by its set of vertices $\{v_1, \ldots, v_n\}$ and its set of edges $\{(v_i, v_{i+1}) \mid 1 \leq i \leq n-1\}$.

\begin{theorem}\label{score_chemin}
Let $P_n$ be a path on $n$ vertices. Then:
\begin{itemize}
    \item If $n$ is even, $Ms(P_n) = Bs(P_n) = 0$.
    \item If $n$ is odd, then $Ms(P_n) = 1$ and $Bs(P_n) = 0$.
\end{itemize}
\end{theorem}

\begin{proof}
\medskip
\begin{itemize}
    \item Suppose first that \( n \) is even.

    Then, Dominator wins the Maker-Breaker domination game on $P_n$. Thus, by Corollary~\ref{neutral}, we have \(\textsc{SHVG}(P_n) = 0\), and consequently, $Ms(P_n) = Bs(P_n) = 0$.
    
    \item Suppose now that $n$ is odd, the outcome of the domination game on $P_n$ is $\mathcal{N}$. So, $Bs(P_n) = 0$, and $Ms(P_n) \ge 1$ by Remark~\ref{dominator}. To prove that $Ms(P_n) \le 1$, set $n = 2k+1$, and denote by $(v_1, \dots, v_{2k+1})$ the vertices of $P_n$, and consider the following pairing strategy for Breaker: if Maker colors a vertex in $\{v_{2i-1}, v_{2i}\}$, for $1 \le i \le k$, Breaker colors the other one. This strategy ensures that no vertex in $v_1, \dots, v_{2k}$ will be happy and thus, the only vertex that Maker can make happy is $v_{2k+1}$, which ensures $Ms(P_n) \le 1$. \qedhere
\end{itemize}
\end{proof}

\subsubsection{Union of paths}

Once the score on a single path can be computed, union of paths can be handled using the group structure provided by Milnor's universe. In particular, if $n$ is even, since, $Ms(P_n) = Bs(P_n) = 0$, $P_n = 0$, and thus for any graph $G$, we have $Ms(G + P_n) = Ms(G)$ and $Bs(G + P_n) = Bs(G)$. 

\begin{lemma}
    Let $n, m$ be two odd integers, we have $Ms(P_n + P_m) = Bs(P_n + P_m) = 1$. Thus $P_n + P_m = 1$.
\end{lemma}

\begin{proof}
    Since $Ms \ge Bs$, it is sufficient to prove $Bs(P_n + P_m) \ge 1$ and $Ms(P_n + P_m) \le 1$.

Suppose Breaker starts. Up to exchange $n$ and $m$, suppose that his first move is a vertex $v_0$ in $P_n$. It's now Maker's turn, and thus Maker can ensure a score of $1$ since $Ms(P_n + P_m, \emptyset, \{v_0\}) \ge Ms(P_m, \emptyset, \emptyset) + Bs(P_n, \emptyset, \{v_0\}) \ge Ms(P_m, \emptyset, \emptyset) =1$ by Theorem~\ref{bound union milnor}.

Suppose now that Maker starts. Let $n = 2k+1$ and $m = 2l +1$. Denote by $u_1, \dots, u_{2k+1}$ the vertices of $P_n$ and by $v_1, \dots, v_{2l+1}$ the vertices of $P_m$. Consider the following pairing for Breaker $\{(u_1, v_1)\} \cup \{(u_{2i}, u_{2i+1})\}_{1 \le i \le k} \cup \{(v_{2i}, v_{2i+1})\}_{1 \le i \le l}$. This strategy ensures that all the vertices $u_i$ and $v_i$ for $i \ge 2$ will not be happy, and that Maker will color at most one of $u_1, v_1$, which ensures that at most one of them will be happy. Thus, $Ms(P_n + P_m) \le 1$.
\end{proof}

\begin{corollary}\label{corollary union of paths}
    Let $k_1, \dots, k_p$ be integers, with $l$ of them being odd. We have $Ms(P_{k_1} + \dots + P_{k_p}) = \left \lceil\frac{l}{2} \right \rceil$ and $Bs(P_{k_1} + \dots + P_{k_p}) = \left \lfloor\frac{l}{2} \right \rfloor$
\end{corollary}

\begin{remark}
    Since cycles have outcome $\mathcal{D}$ in the Maker-Breaker domination game, they can be added to any game without changing the score. Thus, Corollary~\ref{corollary union of paths} can be extended to graphs of maximum degree~$2$.
\end{remark}

\subsection{Subdivided Stars}
Now, we focus on \emph{subdivided stars}. These graphs are formed by connecting a finite number of paths to a central vertex. Subdivided stars represent an intermediate level of complexity between paths and more general tree structures, since they are trees with a single vertex of degree higher than $3$.

\begin{lemma}\label{remove vertex maker}
    Let $(G, M, B)$ be a position of the game. We have for any $v \in M$, $s(G, M, B) \geq
s(G \setminus \{v\}, M \setminus \{v\}, B)$.
\end{lemma}

\begin{proof}
Let $\strat$ be a strategy for Maker in $s(G \setminus \{v\}, M \setminus \{v\}, B)$. By applying $\strat$ in $(G, M, B)$ which has the same set of uncolored vertices, any vertex made happy in $(G, M, B)$ is also happy in $s(G \setminus \{v\}, M \setminus \{v\}, B)$, which proves the result.
\end{proof}

\begin{theorem}\label{subdivided stars}
    Let \(S\) be a subdivided star obtained by connecting $k$ paths $P_1, \dots, P_k$ to a central vertex $C$. If $k \le 2$, $S$ is a path, and the score on $S$ is given by Theorem~\ref{score_chemin}. Otherwise, denote by $l$ the number of odd paths attached to $C$. We have $Ms(S) = \left \lfloor \frac{l}{2} \right \rfloor$ and $Bs(S) = 0$,
    unless $l=0$ in which case $Ms(S)=1$.
\end{theorem}

\begin{proof}
Let $S$ be a subdivided star. First, since the Maker-Breaker domination game on $S$ has outcome $\mathcal{N}$~\cite{duchene2020domination}, we have $Bs(S) = 0$.

Let $C$ be its central vertex and $P_1, \dots, P_k$ be the paths attached to it. Note that if $k \le 2$, $S$ is a path and its score is given by Theorem~\ref{score_chemin}, i.e. its score is $1$ if it has an odd number of vertices and Maker starts, and $0$ otherwise. 

Suppose now that Maker starts and $k > 2$. Let $l$ be the number of odd paths in $S$ and let $P_1, \dots, P_l$ be these $l$ odd-length path. Suppose $l \geq 1$.
If \Maker{} colors the central vertex at the beginning,
by coloring $C$, 
she can guarantee a score of at least $Bs(S, \{C\}, \emptyset)$. By Lemma~\ref{remove vertex maker},  $Bs(S, \{C\}, \emptyset) \geq  Bs(S \setminus C, \emptyset, \emptyset)$ and, since $S \setminus C$ is an union of paths, by Corollary~\ref{corollary union of paths}, $ Bs(S \setminus C, \emptyset, \emptyset) = \lfloor \frac{l}{2} \rfloor$. Hence $Ms(S) \geq \lfloor \frac{l}{2} \rfloor$.


 Let $n_i = 2k_i +1$ be the length of $P_i$ and denote by $u^i_1, \dots, u^i_{2k_i+1}$ the vertices of $P_i$ for $1\leq i \leq l$ such that $u^i_{2k_i+1}$  is connected to $C$. \Breaker{} has a pairing strategy to prevent \Maker{} from scoring inside the even-length paths by pairing adjacent vertices starting from their leaves. Moreover the pairing $ \{(u^1_{2j-1}, u^1_{2j})\}_{1 \le j \le k_1} \cup \{(u^1_{2k_1+1},C)\}$ enables \Breaker{} to prevent \Maker{} from scoring inside $P_1 \cup \{C\}$. It results that the only vertices that can be made happy are in  $ (P_{2} + \dots + P_{l})$. Hence, by applying the strategy provided in Corollary~\ref{corollary union of paths} in $(P_{2} + \dots + P_{l})$ when Maker plays in it, and by applying his pairing strategy in $S \setminus \big (P_{2} + \dots + P_{l}\big)$, since $ Ms(P_{2} + \dots + P_{l}) = \left \lceil\frac{l-1}{2} \right \rceil = \lfloor \frac{l}{2} \rfloor$, we conclude that $Ms(S) \le  \lfloor \frac{l}{2} \rfloor$. All together, we have $Ms(S) =  \lfloor \frac{l}{2} \rfloor$


If $l = 0$ however, \Maker{} can still guarantee a score of $1$. We prove this result by induction. Since $k > 2$, there are at least two even paths $P_1$ and $P_2$. Let $S_n$ be the subdivided star that is formed by connecting $n$ paths $P_1, P_2, \ldots, P_n$ of length exactly $2$ to a central vertex $C$. By coloring successively all the leaves' neighbors, \Maker{} forces \Breaker{} to answer by coloring each leaves, since otherwise by coloring one of the leaves and its neighbor \Maker{} can score $1$, which will conclude the proof.
Once \Maker{} colored all the leaves' neighbors and \Breaker{} colored all the leaves, it is again \Maker{}'s turn. All vertices adjacent to the central vertex have been colored by \Maker{}, thus by coloring it \Maker{} scores $1$. As a result, $Ms(S_n)$ is at least one. This sequence is illustrated in Figure~\ref{fig:strat subdivided star} for $n=3$.

\begin{figure}
    \centering
\begin{center}
    \begin{tikzpicture}[
    scale=0.75,
  var/.style={circle,draw=gray,thick,fill=gray!20,minimum size=8mm},
  varM/.style=
  {circle,draw=red,thick,fill=red!20,minimum size=8mm},
  varB/.style=
  {circle,draw=blue,thick,fill=blue!20,minimum size=8mm},  
  >=latex]
  
  \node[varB] (1) at (0,0) {$2$};
  \node[varM] (2) at (0,-2) {$1$};
  \node[varM] (3) at (0,-4) {$7$};
  \node[varM] (4) at (-2,-4) {$3$};
  \node[varB] (5) at (-4,-4) {$4$};
  \node[varM] (6) at (2,-4) {$5$};
  \node[varB] (7) at (4,-4) {$6$};

  \draw (1) -- (2);
  \draw (2) -- (3);
  \draw (3) -- (4);
  \draw (4) -- (5);
  \draw (3) -- (6);
  \draw (6) -- (7);
    \end{tikzpicture}
\end{center}
    \caption{The strategy on a star with all branches of length $2$. The number on the vertices is the order in which they are played.}
    \label{fig:strat subdivided star}
\end{figure}
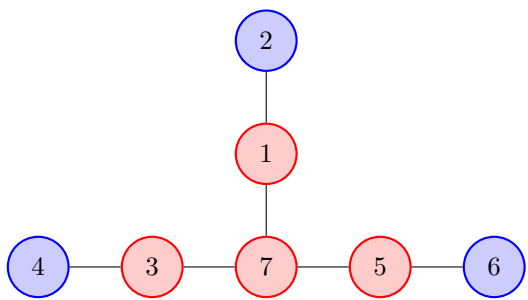

If $S$ is formed by connecting $n$ paths $P_1, P_2, \ldots, P_n$ each of length $2k_i$ for $1 \leq i \leq n$ to a central vertex $C$, and assuming that $P_1$ has length $2k_1>2$. Then \Maker{} colors the neighbor of the leaf of $P_1$, forcing \Breaker{} to answer by playing on the leaf, or letting Maker score $1$ in her next move. Then by Lemma~\ref{remove vertex maker} and Lemma~\ref{equiv decompose}, Maker can ensure on this position a score at least equal to that of $S'$ formed by connecting $n$ paths $P_1', P_2, \ldots, P_n$ to $C$ where $P_1'$ is the path of length $2k_1-2$, and, inductively, at least that of $S_n$. 

For the other direction \Breaker{} has a pairing strategy preventing any vertex besides the central vertex to be happy, by pairing the vertices of each path together, which is possible by again pairing adjacent vertices starting from the leaves. The only vertex that is not paired, and thus that can be happy and the end of the game is $C$, which proves $Ms(S)\le1$. All together, $Ms(S) = 1$.
\end{proof}

\subsection{An \FPT\ algorithm for the neighborhood diversity}\label{sec: parameterized}

The parameterized complexity of positional games has recently grown a strong interest and is a promising way to obtain efficient algorithms for games that are usually hard to solve. Bonnet, Gaspers, Lambilliotte, R\"{u}mmele and Saffidine, have proved that Maker-Breaker positional games are \W[1]-hard parameterized by the number of moves, but can be solved in \FPT\ time for the same parameter in several graph classes~\cite{Bonnet2017}. However, these general results cannot be used when considering scores, and therefore do not apply directly in our framework. In the scoring positional games universe, the first parameter considered in~\cite{bagan2024incidence} is the neighborhood diversity, which was introduced by Lampis~\cite{Lampis2012}, and measures the variety of the neighborhoods of the vertices in the graph. We propose an \FPT-algorithm for this parameter, where $w$ is the neighborhood diversity of the input graph. Note that, since the game is already \NP-hard on caterpillars, one cannot hope, under classical complexity assumptions, for an \FPT-algorithm for most of the general parameters, such as the pathwidth, the treewidth, the twinwidth, or the feedback edge number.

We first recall that two vertices, $u$ and $v$ {\em have the same type} if $N(u) \setminus \{v\} = N(v) \setminus \{u\}$. If $G$ is a graph, the neighborhood diversity of $G$, denoted $nd(G)$ is defined as the smallest integer $w$ such that the vertices of $G$ can be partitioned into $k$ sets $V_1, \dots, V_w$ where all the vertices of each set have the same type. Such a partition is called a {\em neighborhood partition} of $G$. Note that it implies that each set $V_i$ is either a clique or an independent set.

The main idea of the reduction is to use Lemma~\ref{super lemma} to color all the vertices of same type but at most~$1$. However, contrary to the kernelization provided for the game Incidence~\cite{bagan2024incidence}, we cannot here easily remove the claimed vertices, as some of them may require several moves from Maker to contribute to the score. Hence, our \FPT\ algorithm cannot provide us a polynomial kernel, and the question of its existence remains open.

\begin{theorem}\label{neighborhood diversity}
Let \(G = (V,E)\) be a graph of neighborhood diversity \(w\). The score in the \textsc{SHVG} can be computed in \FPT\ time when parameterized by the neighborhood diversity. In particular, \(s(G)\) can be computed in time \(\mathcal{O}(3^w (|E|+|V|) )\).
\end{theorem}

\begin{proof}
We first compute a neighborhood partition of $V$, which can be done in time $O(|E| + |V|)$ time~\cite{Lampis2012}

Let \(u, v\) be two vertices having the same type. By definition, we have \(N(u)\setminus\{v\}=N(v)\setminus\{u\}\). Thus, for any position of the game $(G,M,B)$ and any subset \(X \subseteq V \setminus \{u, v\}\), we have:
\[
\left| \{ w \in V \mid N[w] \subseteq X \cup M \cup \{u\} \} \right| 
= \left| \{ w \in V \mid N[w] \subseteq X \cup M \cup \{v\} \} \right|.
\]

Therefore, Lemma~\ref{super lemma} can be applied to each pair of vertices of the same type.

\medskip

By applying this lemma to all the pairs of uncolored vertices of same type, the resulting position has at most one vertex of each type that remains uncolored and has the same score as the original instance. We can then explore all possible games on these vertices using dynamic programming. Since the number of uncolored vertices is at most $w$, and each vertex can be in one of three states (uncolored, colored by \Maker{}, or colored by \Breaker{}), this exploration can be performed in time \( O(3^w)\).

\medskip

Finally, given a position in which all the vertices are colored, computing the score can be done in \(\mathcal{O}(m + n)\) time. Putting everything together, the overall complexity is \(\mathcal{O}(3^w(|E| + |V|))\).
\end{proof}

\section{Conclusion}

In this paper, we initiated the study of the Maker-Breaker Scoring Happy Vertex Game, which corresponds to a scoring version of the domination game, and we focused on some simple graph classes. Among our considerations, the \NP-hardness proof for caterpillars is related to a question of~\cite{Bagan2025}, which proved that the domination game is in \NP\ when restricted to interval graphs, and asks whether the problem is \NP-complete or not. In the case of the Happy vertex game, the \NP-hardness result for caterpillar makes us wonder whether the game is in \NP\ or \PSPACE-complete for this class of graphs.

Among the other natural classes of graphs to consider, a natural extension of our results for the neighborhood diversity would be to consider graphs of bounded modularwidth, starting naturally with cographs which have modularwidth~$0$ and are known to be polynomial-time solvable in the Maker-Breaker Domination game. However, this study is not as simple here, since isolating a single vertex will not end the game: On playing on a union of cographs having each a single universal vertex, a move from Maker will consist in replacing one of these cographs by the union of its component, while a move of Breaker will consist in removing one of the connected component of the graph. This particular behavior makes it difficult to propose strategies.

Counting the number of winning sets claimed by Maker appears as a natural generalization of a Maker-Breaker game which was first introduced in~\cite{bagan2024incidence}. A natural question, considering instances in which Maker wins, would be to consider other classical positional games on graphs, and to count the number of winning sets she can fill up. The study of the complexity of the other classical positional games on graphs has been studied recently~\cite{DUCHENE2025502}, and observing these games with a scoring view seems to be a natural new step in their study. In particular, the study of the $H$-game led to the study of the number of triangles Maker can create~\cite{GLAZIK2022103536}, and an algorithmic study of this game would be a natural extension of our results.

\bibliography{main}

@article{ZHANG2015,
title = {Algorithmic aspects of homophyly of networks},
journal = {Theoretical Computer Science},
volume = {593},
pages = {117-131},
year = {2015},
issn = {0304-3975},
doi = {https://doi.org/10.1016/j.tcs.2015.06.003},
url = {https://www.sciencedirect.com/science/article/pii/S0304397515005058},
author = {Peng Zhang and Angsheng Li},
}

@inproceedings{Eto2025HappyVertex,
  author    = {Hiroshi Eto and Akiko Fujimoto and Hironori Kiya and Ruka Matsushita and Eiji Miyano and Yuto Murao and Toshiki Saitoh},
  title     = {Happy Vertex Game},
  booktitle = {Proceedings of the International Symposium on Computing and Networking (CANDAR 2025)},
  year      = {2025}
}

@article{duchene2020domination,
  title={Maker--Breaker domination game},
  author={Duchene, Eric and Gledel, Valentin and Parreau, Aline and Renault, Gabriel},
  journal={Discrete Mathematics},
  volume={343},
  number={9},
  pages={111955},
  year={2020},
  publisher={Elsevier}
}

@article{bagan2024incidence,
  title={Incidence, a scoring positional game on graphs},
  author={Bagan, Guillaume and Deschamps, Quentin and Duch{\^e}ne, Eric and Durain, Bastien and Effantin, Brice and Gledel, Valentin and Oijid, Nacim and Parreau, Aline},
  journal={Discrete Mathematics},
  volume={347},
  number={8},
  pages={113570},
  year={2024},
  publisher={Elsevier}
}

@article{lewis2019finding,
  title={Finding happiness: an analysis of the maximum happy vertices problem},
  author={Lewis, Rhyd and Thiruvady, Dhananjay and Morgan, Kerri},
  journal={Computers \& Operations Research},
  volume={103},
  pages={265--276},
  year={2019},
  publisher={Elsevier}
}

@article{zhang2018improved,
  title={Improved approximation algorithms for the maximum happy vertices and edges problems},
  author={Zhang, Peng and Xu, Yao and Jiang, Tao and Li, Angsheng and Lin, Guohui and Miyano, Eiji},
  journal={Algorithmica},
  volume={80},
  number={5},
  pages={1412--1438},
  year={2018},
  publisher={Springer}
}

@inproceedings{agrawal2017parameterized,
  title={On the parameterized complexity of happy vertex coloring},
  author={Agrawal, Akanksha},
  booktitle={International Workshop on Combinatorial Algorithms},
  pages={103--115},
  year={2017},
  organization={Springer}
}

@inproceedings{aravind2016linear,
  title={Linear time algorithms for happy vertex coloring problems for trees},
  author={Aravind, NR and Kalyanasundaram, Subrahmanyam and Kare, Anjeneya Swami},
  booktitle={International Workshop on Combinatorial Algorithms},
  pages={281--292},
  year={2016},
  organization={Springer}
}

@article {Dokyeesun2025,
author = {Dokyeesun, Pakanun},
title = {Maker-Breaker domination game on Cartesian products of graphs},
journal = {Communications in Combinatorics and Optimization},
volume = {},
number = {},
pages = {-},
year  = {2025},
publisher = {Azarbaijan Shahid Madani University},
issn = {2538-2128}, 
eissn = {2538-2136}, 
doi = {10.22049/cco.2025.29866.2198},	
url = {https://comb-opt.azaruniv.ac.ir/article_14917.html},
eprint = {https://comb-opt.azaruniv.ac.ir/article_14917_7912edad7574bc7abc41f3b736b7f1c9.pdf}
}

@article{Bagan2025,
  author  = {Bagan, Guillaume and Duch{\^e}ne, {\'E}ric and Gledel, Valentin and Lehtil{\"a}, Tuomo and Parreau, Aline},
  title   = {Partition Strategies for the Maker--Breaker Domination Game},
  journal = {Algorithmica},
  year    = {2025},
  volume  = {87},
  number  = {2},
  pages   = {191--222},
  doi     = {10.1007/s00453-024-01280-x},
  url     = {https://doi.org/10.1007/s00453-024-01280-x},
  issn    = {1432-0541},
}

@InProceedings{oijid2025qbf,
author="Oijid, Nacim",
editor="Finocchi, Irene
and Georgiadis, Loukas",
title="Bounded Degree QBF and Positional Games",
booktitle="Algorithms and Complexity",
year="2025",
publisher="Springer Nature Switzerland",
address="Cham",
pages="121--135",
isbn="978-3-031-92935-9"
}

@incollection{hales1963regularity,
  title={Regularity and positional games},
  author={Hales, Alfred W and Jewett, Robert I},
  booktitle={Classic Papers in Combinatorics},
  pages={320--327},
  year={2009},
  publisher={Springer}
}

@article{schaefer1978complexity,
  title={On the complexity of some two-person perfect-information games},
  author={Schaefer, Thomas J},
  journal={Journal of computer and system Sciences},
  volume={16},
  number={2},
  pages={185--225},
  year={1978},
  publisher={Elsevier}
}

@article{rahman20236,
  title={6-uniform Maker-Breaker game is PSPACE-complete},
  author={Rahman, Md Lutfar and Watson, Thomas},
  journal={Combinatorica},
  volume={43},
  number={3},
  pages={595--612},
  year={2023},
  publisher={Springer}
}

@article{koepke2025advances,
  title={Solving Maker-Breaker Games on 5-uniform hypergraphs is PSPACE-complete},
  author={Koepke, Finn Orson},
  journal={arXiv preprint arXiv:2502.20271},
  year={2025}
}

@misc{galliot20254uniformmakerbreakermakermakergames,
      title={4-uniform Maker-Breaker and Maker-Maker games are PSPACE-complete}, 
      author={Florian Galliot},
      year={2025},
      eprint={2509.13819},
      archivePrefix={arXiv},
      primaryClass={cs.DM},
      url={https://arxiv.org/abs/2509.13819}, 
}

@misc{galliot2025makerbreakersolvedpolynomialtime,
      title={Maker-Breaker is solved in polynomial time on hypergraphs of rank 3}, 
      author={Florian Galliot and Sylvain Gravier and Isabelle Sivignon},
      year={2025},
      eprint={2209.12819},
      archivePrefix={arXiv},
      primaryClass={cs.DM},
      url={https://arxiv.org/abs/2209.12819}, 
}

@inproceedings{berman2002some,
  title={On some tighter inapproximability results},
  author={Berman, Piotr and Karpinski, Marek},
  booktitle={Automata, Languages and Programming: 26th International Colloquium, ICALP’99 Prague, Czech Republic, July 11--15, 1999 Proceedings},
  pages={200--209},
  year={2002},
  organization={Springer}
}

@book{larsson2019games,
  title={Games of No Chance 5},
  author={Larsson, Urban},
  volume={5},
  year={2019},
  publisher={Cambridge University Press}
}

@article{milnor1953sums,
  title={Sums of positional games},
  author={Milnor, John},
  journal={Ann. of Math. Stud.(Contributions to the Theory of Games, HW Kuhn and AW Tucker, eds.), Princeton},
  volume={2},
  number={28},
  pages={291--301},
  year={1953}
}

@phdthesis{oijid2024thesis,
  title={Complexit{\'e} des jeux positionnels sur les graphes},
  author={Oijid, Nacim},
  year={2024},
  school={Universit{\'e} Claude Bernard-Lyon I}
}

@InProceedings{Bonnet2017,
  author =	{Bonnet, \'{E}douard and Gaspers, Serge and Lambilliotte, Antonin and R\"{u}mmele, Stefan and Saffidine, Abdallah},
  title =	{{The Parameterized Complexity of Positional Games}},
  booktitle =	{44th International Colloquium on Automata, Languages, and Programming (ICALP 2017)},
  pages =	{90:1--90:14},
  series =	{Leibniz International Proceedings in Informatics (LIPIcs)},
  ISBN =	{978-3-95977-041-5},
  ISSN =	{1868-8969},
  year =	{2017},
  volume =	{80},
  editor =	{Chatzigiannakis, Ioannis and Indyk, Piotr and Kuhn, Fabian and Muscholl, Anca},
  publisher =	{Schloss Dagstuhl -- Leibniz-Zentrum f{\"u}r Informatik},
  address =	{Dagstuhl, Germany},
  URL =		{https://drops.dagstuhl.de/entities/document/10.4230/LIPIcs.ICALP.2017.90},
  URN =		{urn:nbn:de:0030-drops-74941},
  doi =		{10.4230/LIPIcs.ICALP.2017.90}
}

@article{Lampis2012,
  author  = {Lampis, Michael},
  title   = {Algorithmic Meta-theorems for Restrictions of Treewidth},
  journal = {Algorithmica},
  volume  = {64},
  number  = {1},
  pages   = {19--37},
  year    = {2012},
  doi     = {10.1007/s00453-011-9554-x},
  url     = {https://doi.org/10.1007/s00453-011-9554-x},
}

@article{DUCHENE2025502,
title = {Complexity of Maker–Breaker games on edge sets of graphs},
journal = {Discrete Applied Mathematics},
volume = {361},
pages = {502-522},
year = {2025},
issn = {0166-218X},
doi = {https://doi.org/10.1016/j.dam.2024.11.012},
url = {https://www.sciencedirect.com/science/article/pii/S0166218X24004785},
author = {Eric Duchêne and Valentin Gledel and Fionn {Mc Inerney} and Nicolas Nisse and Nacim Oijid and Aline Parreau and Miloš Stojaković},
}

@article{GLAZIK2022103536,
title = {A new bound for the Maker–Breaker triangle game},
journal = {European Journal of Combinatorics},
volume = {104},
pages = {103536},
year = {2022},
issn = {0195-6698},
doi = {https://doi.org/10.1016/j.ejc.2022.103536},
url = {https://www.sciencedirect.com/science/article/pii/S0195669822000324},
author = {Christian Glazik and Anand Srivastav},
}

@article{Hanner1959MeanPO,
  title={Mean play of sums of positional games},
  author={Olof Hanner},
  journal={Pacific Journal of Mathematics},
  year={1959},
  volume={9},
  pages={81-99},
  url={https://api.semanticscholar.org/CorpusID:120425420}
}

@article{DUCHENE2024114274,
title = {Bipartite instances of INFLUENCE},
journal = {Theoretical Computer Science},
volume = {982},
pages = {114274},
year = {2024},
issn = {0304-3975},
doi = {https://doi.org/10.1016/j.tcs.2023.114274},
url = {https://www.sciencedirect.com/science/article/pii/S030439752300587X},
author = {Eric Duchêne and Nacim Oijid and Aline Parreau},
}

@inproceedings{stockmeyer1973,
  title={Word problems requiring exponential time (preliminary report)},
  author={Stockmeyer, Larry J. and Meyer, Albert R.},
  booktitle={Proceedings of the fifth annual ACM symposium on Theory of computing},
  pages={1--9},
  year={1973}
}

\end{document}